\documentclass[runningheads]{llncs}
\usepackage[T1]{fontenc}
\usepackage{url}
\usepackage{hyperref}
\usepackage{listing}
\usepackage{graphicx}
\usepackage{xspace}
\usepackage{xcolor}
\usepackage{listings}
\usepackage{orcidlink}

\usepackage[frozencache=true]{minted}

\usepackage[misc]{ifsym} 

\usepackage{graphicx} 
\usepackage[firstpage]{draftwatermark} 

\newcommand\haclstar{HACL$^\star$\xspace}

\newcommand*{\li}{\lstinline[keywordstyle=\ttfamily\color{black},basicstyle=\ttfamily\color{black},commentstyle=\ttfamily\color{black}]}

\newcommand{\comments}{true}
\usepackage{ifthen}
\ifthenelse{\equal{\comments}{true}}
{
  \DeclareRobustCommand{\son}[1]{ {\begingroup\color{red!60!black}{(Son) #1}\endgroup} }
  \DeclareRobustCommand{\af}[1]{ {\begingroup\color{green!60!black}{(Aymeric) #1}\endgroup} }
  \DeclareRobustCommand{\jonathan}[1]{ {\begingroup\color{blue!60!black}{(Jonathan) #1}\endgroup} }
  \DeclareRobustCommand{\gb}[1]{ {\begingroup\color{teal!50!green}{(Guillaume) #1}\endgroup} }
}
{
  \DeclareRobustCommand{\son}[1]{ {} }
  \DeclareRobustCommand{\af}[1]{ {} }
  \DeclareRobustCommand{\jonathan}[1]{ {} }
  \DeclareRobustCommand{\gb}[1]{ {} }
}

\newcommand{\sref}[1]{\S\ref{sec:#1}}
\newcommand{\fref}[1]{Figure~\ref{fig:#1}}

\begin{document}
\title{Charon: An Analysis Framework for Rust}


\author{Son Ho\inst{1}\textsuperscript{\Letter}\orcidlink{0000-0003-3297-9156} \and
 Guillaume Boisseau\inst{2}\orcidlink{0000-0001-5244-893X} \and
 Lucas Franceschino\inst{3} \and
 Yoann Prak\inst{2} \and
 Aymeric Fromherz\inst{2}\orcidlink{0000-0003-2642-543X} \and
 Jonathan Protzenko\inst{4}\orcidlink{0000-0001-7347-3050}
}

\institute{
 Microsoft Azure Research, UK,
 \textsf{t-sonho@microsoft.com},
 \and Inria Paris, France,\\\textsf{\{guillaume.boisseau,yoann.prak,aymeric.fromherz\}@inria.fr},
 \and Cryspen, France, \textsf{lucas@cryspen.com},
 \and Microsoft Azure Research, USA, \textsf{protz@microsoft.com}
}
\authorrunning{S. Ho, G. Boisseau, L. Franceschino, Y. Prak, A. Fromherz, and J. Protzenko}
%
\maketitle              
\begin{abstract}
    With the explosion in popularity of the Rust programming language, a wealth
  of tools have recently been developed to analyze, verify, and test Rust
  programs.  Alas, the Rust ecosystem remains relatively young, meaning that
  every one of these tools has had to re-implement difficult, time-consuming
  machinery to interface with the Rust compiler and its cargo build system, to
  hook into the Rust compiler's internal representation, and to expose an
  abstract syntax tree (AST) that is suitable for analysis rather than optimized
  for efficiency.

  We address this missing building block of the Rust ecosystem, and propose
  Charon, an analysis framework for Rust. Charon acts as a swiss-army knife for
  analyzing Rust programs, and deals with all of the tedium above, providing
  clients with an AST that can serve as the foundation of many
  analyses. We demonstrate the usefulness of Charon through a series of case
  studies, ranging from a Rust verification framework (Aeneas), a compiler
  from Rust to C (Eurydice), and a novel taint-checker for cryptographic code.
  To drive the point home, we also re-implement a popular existing analysis
  (Rudra), and show that it can be replicated by
  leveraging the Charon framework.
\end{abstract}

\section{Introduction}


Over the past decade, the Rust programming language has gained traction
both in academia and in
industry~\cite{creusot,prusti,refinedrust,yuga,rudra,verus,microsoft2022rust},
consistently ranking as the most beloved language by
developers~\cite{github2023rust,stackoverflow2023rust} for the past 8 years. A
large part of this success stems from several key features of the language: Rust
provides both the high
performance and low-level idioms commonly associated to C or C++, as well as
memory-safety by default thanks to its rich, borrow-based type system. This
makes Rust suitable for a wide range of applications: both
Windows~\cite{rustinwindows} and the Linux kernel~\cite{rustlinux} now support
Rust, the latter marking the first time a language beyond C was ever approved
for Linux. The safety guarantees of Rust are particularly appealing
for security-critical systems: leading governments now also recommend
Rust~\cite{whitehouse2024,nsa2022safelang}.

Of course, despite being safer than C or C++, Rust programs are not immune
to bugs and vulnerabilities. Case in point, between January 1, 2024 and January
1, 2025, 137 security advisories against Rust crates
were filed on RUSTSEC, a vulnerability database for the Rust ecosystem
maintained by the Rust Secure Code working group. These vulnerabilities arose
due to several reasons: runtime errors leading to aborted executions (\li`panic`
in Rust, e.g., after an out-of-bounds array access),
implementation or design flaws, or even memory vulnerabilities when using Rust's \li`unsafe`
escape hatch, which allows the use of C-like, unchecked pointer operations when Rust's borrow-based
type system is too restrictive.

To enforce those properties that fall outside the scope of Rust's
borrow-checker, a vibrant ecosystem of static analyzers~\cite{yuga,rudra,miri}, model
checkers~\cite{kani,stateright} and deductive verification
tools~\cite{hacspec,creusot,verus,rustbelt,prusti,gillian-rust,refinedrust,ho2022aeneas}
has been proposed to reason about and analyze Rust programs.
However, developing a new tool targeting Rust currently requires an important
engineering effort to meaningfully and efficiently interact with the
Rust compiler, \li`rustc`.
First, Rust is a complex language, and just like many C analyses plug into \li`libclang`,
Rust analyses need to plug into \li`rustc`: reimplementing a Rust frontend from scratch
would be a significant undertaking.
Unfortunately, the various intermediate representations (IRs) in \li+rustc+ are
optimized for speed and efficiency, not ease of consumption by analysis tools.
Specifically, rather than provide a fully decorated abstract syntax tree (AST)
with all available information, \li`rustc` instead
exposes \emph{queries} to, e.g., obtain the type of an expression only when
needed. While this design allows for efficient incremental
compilation, it makes life harder for tool authors, since they have to deal with
additional levels of indirection, and information scattered across multiple
global tables and IRs; this style of APIs requires deep
knowledge of the compiler internals.
Additionally, while sufficient for compilation, the information provided by the compiler
sometimes needs to be expanded for analysis purposes.
For instance, querying the trait solver only gives partial information about the trait
instances used at a point in the code.
Reconstructing this information requires additional, error-prone work from tool authors.
%
%
Finally, the Rust compiler itself is not invoked in isolation; any non-trivial
Rust project uses the Cargo build system, which collects dependencies,
synthesizes \li+rustc+ invocations with suitable library and include paths, and
generally drives \li+rustc+. A realistic analysis tool must therefore hook
itself onto Cargo, a non-trivial endeavor that again requires deep knowledge of
Cargo and \li+rustc+.

To address these limitations, we present Charon, a Rust analysis
framework providing an analysis-oriented interface to the Rust
compiler and tooling, which allows tool authors to focus on the core of their analysis,
rather than on mundane details of the Rust compiler internals.
%
%
Our contributions are as follows.
Through the development of a variety of tools atop \li+rustc+,
we are of the opinion that many of MIR's design choices are unsuitable for analysis,
such as: low-level pattern-matching that switches over the enumeration
tag; encoding bounds-checks semantics with assertions; a precompiled
representation of constants as byte arrays; and many more.
Our first contribution is thus the design of ULLBC (Unstructured Low-Level Borrow Calculus)
and LLBC (Low-Level Borrow Calculus), two dual views over Rust's
internals that offer a control-flow graph (CFG) and an AST respectively. Both preserve the low-level MIR semantics
of moves, copies, and explicit borrows and reborrows, but reconstruct constants, shallow pattern matches, and
checked operations, so as to provide a semantically simple view of MIR that is
suitable for further analyses. LLBC is obtained from ULLBC via a Relooper-like
control-flow reconstruction~\cite{ramsey2022beyond,peterson1973capabilities}.
Our second contribution is the engineering of Charon itself, and the accompanying
ecosystem integration. We write a standalone, reusable
infrastructure that allows a client to integrate cleanly with Cargo's build
system. Furthermore, to expose clean ULLBC and LLBC representations (above),
we hide away the complexity of querying the Rust compiler internals, thus offering
\emph{usable} APIs without the incidental complexity.
%
%
Our third and final contribution is a series of case studies that not
only informed the design of LLBC and ULLBC, but also served as an
experimental validation of their wide-ranging applicability. We thus provide empirical
evidence that the design choices
of Charon are suitable for a wide variety of use-cases, supporting our claim
that Charon is the swiss-army knife of Rust analysis.
%

\paragraph{Data-Availability.} Charon is developed publicly on Github~\cite{charon-github}
and released under an open-source license; to foster reproducibility,
an artifact is available online~\cite{charon-artifact}.

\section{The Charon Framework}

We now describe the architecture of both the Rust compiler and the Charon
framework; \fref{picasso} recaps the various steps in visual form.

\begin{figure}[h]
  \centering
  \includegraphics[width=\textwidth]{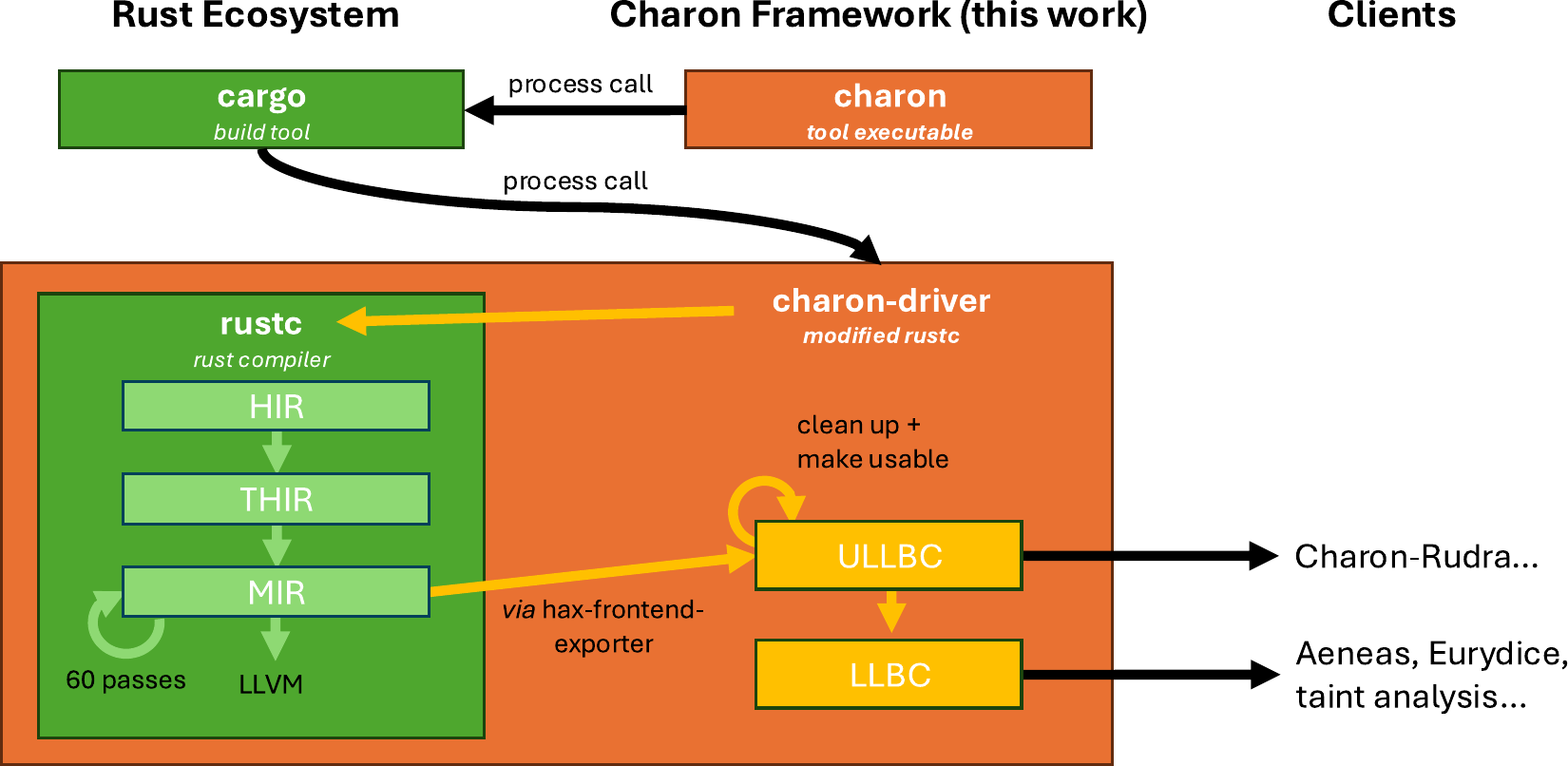}
  \caption{Architectural diagram of the Charon framework, its relationship to
  the Rust compiler, and to consumers.}
  \label{fig:picasso}
  \vspace{-3em}
\end{figure}

\subsection{Background: the Rust Compiler}
The Rust compiler pipeline relies on three ASTs after initial parsing: HIR (``High-level IR''), THIR
(``Typed HIR''), and MIR (``Mid-level IR'').
HIR is the result of expanding macros and resolving names.
Then, in THIR, all type information is
filled, and a first round of desugaring is performed, notably for
reborrows, as well as
automatic borrowing and dereferencing.
At this stage, many fine points of semantics are still implicit: moves, copies,
drops, control-flow of patterns, and many more, do not appear in this AST.
MIR is where all these semantic details are made explicit.
MIR is a CFG with a limited set of statements and terminators,
making the semantics lower-level than in HIR and THIR.
%
The Rust compiler features nearly 60 compilation
passes~\cite{mirpasses}
operating on the MIR AST -- a standard design choice, where in practice
several phases rely on different subsets of MIR.
%
The final step, after all optimizations have been run on MIR, consists in
emitting LLVM bitcode, then handing off the rest of the compilation pipeline to
LLVM itself.
We now review our design choices, and explain how they facilitate the task
of developing analyses for Rust.

\subsection{Charon Overview}

As MIR explicits many fine points of semantics that make Rust hard to accurately model,
it is a common starting point for verification tools~\cite{creusot,kani,rudra,yuga2023},
as well as compiler analyses such as borrow-checking, which were too error-prone on
THIR~\cite{thirbugs,mirBlog}.
In particular, moves, copies, (re)borrows, and drops, the core of Rust's semantics,
are all explicit in MIR: for instance, \li+let x = &mut y; f(x)+ becomes, in MIR,
\li+let x = &mut y; f(move &mut(*x))+, so as to avoid invalidating \li+x+ itself via a move-out. Other desugarings,
e.g., for pattern-matches, which have a highly non-trivial semantics in THIR, also appear explicitly in MIR only.
To allow a precise analysis of Rust programs, Charon therefore also operates on MIR.

\paragraph{ULLBC.}
From the MIR representation, Charon constructs a cleaned-up, decorated view called ULLBC.
ULLBC is, like MIR, a CFG; but unlike MIR, ULLBC offers immediate contextual and semantic information
(\sref{reconstruct}), and hides implementation-specific details (\sref{reprs}), while nevertheless
exposing the entire Rust language. In particular, rather than have the user query
\li`rustc`, e.g.,
to get type information about auxiliary data structures or to invoke the trait solver, ULLBC has
all of this critical information directly attached to the CFG.

Additionally, ULLBC features several clean-up transformation passes that offer a more structured, semantic
view of MIR.
These passes, e.g., simplify pattern-matching operating on low-level representation to replace them with high-level, ML-style pattern matches;
transform a variety of desugarings of \li`panic!` to a single, unified statement, or pack dynamic checks for integer overflows
with their corresponding arithmetic operation to simplify the semantics.
We envision that consumers of ULLBC ultimately will decide whether these passes are fit for them; for instance,
specific consumers might want to see explicit assertions for checked operations. Should they do so, we envision
a general API where they can manually choose which reconstruction passes to enable.

\paragraph{LLBC.}
Charon then applies a control-flow reconstruction algorithm on a ULLBC CFG to generate an LLBC AST that materializes
structured loops and branching. In practice, we reverse-engineer \li`rustc`'s CFG construction, and recreate loops,
conditionals, etc. out of the MIR basic blocks.

\subsection{Limitations}

Charon already supports a sizable subset of Rust, as evidenced by our evaluation
(\sref{case-studies}).
However some less-central features are not supported, which consist of:
dynamic trait dispatch (\li+dyn Trait+), generic associated types, trait aliases, async,
and  \li+impl+s in the return type of functions.
To support analyzing crates that use such features,
Charon was designed to be robust: a declaration that cannot be translated is marked as missing
and the rest of the crate is translated as usual, resulting in a well-formed (if incomplete)
(U)LLBC. So far, none of these missing features have impeded our
case studies.

\subsection{Reconstructing Compiler Information}
\label{sec:reconstruct}
When operating on \li`rustc`, retrieving information needed for program analysis is tricky for two reasons.
First, instead of exposing a fully-decorated representation, \li`rustc` relies on (poorly documented) lazy
computations to \emph{query} the compiler internals. One thus has to skim through the code of \li`rustc` itself
to find which auxiliary function, given a program node, may retrieve the desired information.
Second, and more importantly, some information can be partial or missing.

As an example, consider a Rust trait method call; to analyze it, we first need the instance of the
corresponding Rust \emph{trait} with
its proper arguments (i.e., the instantiated \emph{impl}), and the reference to the method including its arguments
(in case the method is generic). Unfortunately, the concrete trait instance is not
directly available in MIR.
For instance, consider the snippet of code below.

\begin{minted}{rust}
fn clone_vec<T>(v : &Vec<T>) -> Vec<T> where T:Clone { v.clone() }
\end{minted}

When compiling this code, \li+rustc+ needs to assess that it is legal to call the \li+clone+
method by querying the trait solver to find an instance of \li+Clone+ for \li+Vec<T>+;
in our case, this instance is the standard library \li+Clone+ implementation for
\li+Vec<T : Clone>+ (which we will call \li+CloneVec+),
composed with the local instance \li+T : Clone+ received as input
(which we will call \li+CloneT+).

The trait solver only needs to assess that \emph{there
exists} such an instance.
However, when analyzing this code, we might want more precise information, namely that
\li+v.clone()+ uses exactly the implementation \li+CloneVec+ composed with \li+CloneT+.
Unfortunately, retrieving this information from \li+rustc+ requires several non-trivial steps.
First, we need to query the trait solver; to do so, one needs to handle several complex
\li`rustc` concepts, such as its internal representation for binders or late-bound and
early-bound regions.
Worse, the information given by the trait solver is partial.
The solver only tells us that \li+v.clone()+ uses \li+CloneVec+ composed with
\emph{some} instance of \li+T : Clone+ which can be derived from the local \li+where+
clauses of the function, without exhibiting this derivation.
Reconstructing it is a non-trivial problem in general especially as some trait
instances can be implied by other trait clauses in the presence of super traits.\\

Another issue stems from the input parameters given to function and method calls.
Consider the following snippet of code.

\begin{minted}{rust}
fn f<V>(x : V);
fn g<V>(x : V) { f::<V>(x); }
trait Trait<U> { fn f<V>(x : V); }
fn h<T,U,V>(x:V) where T:Trait<U> { (T as Trait<U>)::f::<V>(x) }
\end{minted}

When retrieving the list of generic parameters received by the call to the \emph{function}
\li+f+ in \li+g+, \li+rustc+ gives us \li+V+. However, when querying the parameters
given to the \emph{method} call \li+Trait::f+ in \li+h+, \li+rustc+ concatenates the parameters of
the trait instance with the parameters given to the method itself, yielding \li+T+ (for the
implicit \li+Self+ parameter), \li+U+, and \li+V+,
where one only expects \li+V+. This confusing behavior is error-prone; as a consequence
Charon retrieves and truncates the list of parameters through the following boilerplate
code, which itself relies on several (omitted) auxiliary helpers that we introduced for this purpose
(in red).

\begin{small}
\definecolor{MyRed}{HTML}{800000}
\newcommand*{\boldcode}{\lstinline[keywordstyle=\ttfamily\color{MyRed},basicstyle=\ttfamily\color{MyRed}]}
\begin{minted}[escapeinside=??]{rust}
let (gens, source) = // Is this a function or method call?
 if let Some(assoc) = tcx.opt_associated_item(id) { ... // Omitted
  match assoc.container { // Trait "decl" or impl method call?
   AssocItemContainer::TraitContainer => { // Trait "decl" method call
    let num_cont_gens = tcx.?\boldcode{generics_of(cont_id).own_params}?.len();
    let impl_expr = ?\boldcode{self_clause_for_item}?(s, &assoc, gens).unwrap();
    let method_gens = &gens[num_cont_gens..];
    (method_gens.sinto(s), Some(impl_expr)) }
   AssocItemContainer::ImplContainer => { // Trait impl method call
    let cont_gens = tcx.?\boldcode{generics_of(cont_id)}?;
    let cont_gens = gens.?\boldcode{truncate_to(tcx, cont_gens)}?;
    let mut comb_trait_refs = ?\boldcode{solve_req_traits(s, cont_id, cont_gens)}?;
    comb_trait_refs.extend(std::mem::take(&mut trait_refs));
    trait_refs = comb_trait_refs;
    (gens.sinto(s), None) } } } else {(gens.sinto(s), None)}; // Fun call
\end{minted}
\end{small}

There are many other technicalities to consider leading to boilerplate code;
for instance retrieving the trait instances alone requires more than 600 lines of code.
In contrast, Charon provides the following LLBC
datatype, which directly exposes
all relevant trait information with the properly truncated list of parameters.
The information includes support for type and trait polymorphism,
at the heart of Rust's generic programming facilities, whose
representation relies on a novel, abstract language of trait clauses, parent clauses, self
types and trait bounds.
For instance, a trait might refer to a top-level implementation but also a local where
clause (e.g., \li+T : Clone+), or a trait implied by an associated type (e.g.,
\li+IntoIterator+ contains an associated type \li+IntoIter : Iterator<...>+).

\begin{minted}{rust}
struct FnPtr {
 func: FunIdOrTraitMethodRef, // Function identifier
 generics: GenericArgs, }     // Generic parameters

enum FunIdOrTraitMethodRef {
 Fun(FunId),                            // Top-level function
 Trait(TraitRef, TraitItemName, ...), } // Trait method

struct TraitRef { kind: TraitRefKind, ... }
enum TraitRefKind {
 TraitImpl(TraitImplId, GenericArgs), // Top-level impl
 Clause(ClauseId),                    // Local where clause
 ItemClause(Box<TraitRefKind>, ...),  // Implied by assoc. type
\end{minted}

Beyond trait resolution, Charon also reconstructs the control-flow by applying a
Relooper-like algorithm, packages crates into easy-to-use structures,
and optionally lifts trait associated types into type parameters
while normalizing types which are known to be equal (because, e.g., of a clause
\li+T::Item = u32+).

\subsection{Simplifying Representations}
\label{sec:reprs}

To efficiently compile Rust projects, \li`rustc` stores program information in a variety of representations,
sometimes too low-level to be directly usable for analysis purposes.
In particular, in MIR, constants are already compiled, meaning that instead of a struct
value, one may be simply provided with an array of bytes already laid out.
For instance, let us describe the process of retrieving the high-level representation
of an enumeration value which got compiled to a constant.
We start from a \li+mir::Const+ enumeration, over which we match, retrieving either
a \li+ty::Const+ if the constant is used in a type (e.g., it is used to instantiate a
const generic), or a \li+ConstValue+ if it doesn't.
Diving further, from a \li+ty::Const+, in \emph{one} case (there are many other cases) we get
a \li+ValTree+; by using the type of the constant we learn that, as it is an abstract data type (ADT), we
should call a specific \li+rustc+ helper to turn it into a \li+DestructuredConst+ (below), from which
we can recursively reconstruct our enumeration value.

\begin{minted}{rust}
struct DestructuredConst<'tcx> {
  variant: Option<VariantIdx>,
  fields: &'tcx [ty::Const<'tcx>], }
\end{minted}

The other cases are similarly complex, with many corner cases.
In constrast, Charon provides the following unified, high-level representation for constants.

\begin{minted}{rust}
enum ConstantKind {
  Adt(Option<VariantId>, Vec<Constant>), ... /* Omitted */ }
struct Constant { value: ConstantKind, ty: Ty, }
\end{minted}


Beyond constants, Charon also transforms other low-level representations into high-level,
functional datatypes. This includes, e.g., simplifying the representation of names,
reconstructing span information, removing the uses of the \li+Steal+ datastructure which
allows \li+rustc+ to update definitions in-place through the compilation process at the cost
of making some definitions unavailable if they are accessed in the wrong order,
clarifying the use of binders by using an explicit and uniform treatment of parameters
rather than, e.g., what \li+rustc+ dubs "early bound" and "late bound" region variables,
or simplifying trait implementations by turning default methods into regular methods.

\subsection{A Primer of LLBC}

In the previous sections we presented the transformations performed by Charon to make the MIR
easier to consume; let us now have a quick overview of the output of Charon, focusing on
LLBC.
In LLBC, a crate (\li+TranslatedCrate+, below) contains a crate name, the content of
the source files, and the declarations, where each declaration is uniquely identified
(with an id of type, e.g., \li+TypeDeclId+).
We also compute the groups of mutually recursive declarations (not shown here) that we
order topologically, as it is useful both for analysis and compilation purposes.

\begin{minted}{rust}
struct TranslatedCrate {
  crate_name: String, files: Vector<FileId, File>,
  type_decls: Vector<TypeDeclId, TypeDecl>,
  fun_decls: Vector<FunDeclId, FunDecl>, ... /* omitted */ }
\end{minted}

Function declarations and signatures are straightforward.
Function declarations contain a unique identifier,
metadata for the declaration name, the span and the attributes,
a signature, and an optional body which is omitted if Charon encountered an error
or if the function was marked with the \li+#[charon::opaque]+ attribute.
Signatures simply contain the generic parameters, the list of input types, and the output
type.

\smallskip
\begin{minipage}[t]{0.50\textwidth}
\begin{minted}{rust}
struct FunDecl {
 id: FunDeclId, meta: ItemMeta,
 signature: FunSig,
 body: Result<Body, Opaque>,
 ... /* omitted */ }
\end{minted}
\end{minipage}
\begin{minipage}[t]{0.50\textwidth}
\begin{minted}{rust}
struct FunSig {
 ... /* omitted */,
 generics: GenericParams,
 inputs: Vec<Ty>,
 output: Ty, }
\end{minted}
\end{minipage}

\smallskip

Importantly, the generic parameters (\li+GenericParams+) contain all the information
related to the bound variables (region variables, type variables, etc.) and
the \li+where+ clauses in a simple and explicit format.
This type gathers in one place information which otherwise requires querying \li+rustc+ several
times, and is uniformly used by all the definitions (functions, types, traits
declarations, etc.).

\begin{minted}{rust}
struct GenericParams {
 regions: Vector<RegionId, RegionVar>,
 types: Vector<TypeVarId, TypeVar>,
 const_generics: Vector<ConstGenericVarId, ConstGenericVar>,
 trait_clauses: Vector<TraitClauseId, TraitClause>,  // T : Clone
 regions_outlive: Vec<RegionBinder<RegionOutlives>>, // 'a : 'b
 types_outlive: Vec<RegionBinder<TypeOutlives>>,     // T : 'a
 trait_type_constraints: ..., }                  // T::Item = u32
\end{minted}

The other definitions for types, traits, etc., all follow a similar model by containing a
unique identifier, metadata (i.e., \li+ItemMeta+), generic parameters and an optional
body; we omit them here.
Skipping the definition of function bodies, which store the list of local variables
as well as a block of statements, let us now look at the definition of LLBC statements.
Statements have the expected kinds such as: assignments, function calls, returns, or loops.
The \li+Switch+ enumeration (not shown here) distinguishes \li+if then else+s, switches over
integers, and matches over enumerations. We also preserve code spans and user comments.

\smallskip
\begin{minipage}[t]{0.50\textwidth}
\begin{minted}{rust}
enum StatementKind {
 Assign(Place, Rvalue),
 Call(Call),
 Abort(AbortKind), // panic
 Switch(Switch), Loop(Block),
 Return, Nop, Drop(Place),
 Break(usize), Continue(usize),
 ... /* omitted */ }
\end{minted}
\end{minipage}
\begin{minipage}[t]{0.50\textwidth}
\begin{minted}{rust}
struct Statement {
 span: Span,
 kind: StatementKind,
 comments: Vec<String>, }

struct Block {
 span: Span,
 statements: Vec<Statement>, }
\end{minted}
\end{minipage}
\smallskip

We finish this quick overview with function calls. We carefully wrote the \li+Call+ structure
so that all cases are grouped in one definition and are easy to distinguish. The important
field is \li+FnOperand+, which covers the different cases, that is: the use of
top-level functions, of trait methods, and of function pointers stored in local
variables (e.g., because of the use of a anonymous functions).

\smallskip
\begin{minipage}[t]{0.35\textwidth}
\begin{minted}{rust}
struct Call {
 func: FnOperand,
 args: Vec<Operand>,
 dest: Place, }

enum FnOperand {
 Regular(FnPtr),
 Move(Place), }
\end{minted}
\end{minipage}
\begin{minipage}[t]{0.65\textwidth}
\begin{minted}{rust}
struct FnPtr {
 func: FunIdOrTraitMethodRef,
 generics: GenericArgs, }

enum FunIdOrTraitMethodRef {
 Fun(FunId), // top-level function
 TraitMethod(...), // trait method }
\end{minted}
\end{minipage}
\smallskip

When \li+FnOperand+ refers to a top-level function or a trait method, it uses a
\li+FnPtr+ to bundle a function identifier with its generic arguments. When the \li+FnPtr+
itself identifies a
top-level function, it directly refers to its unique identifier, that we can use to,
e.g., lookup the function definition from the \li+TranslatedCrate+ shown above.
When the \li+FnPtr+ identifies a trait method call, it bundles a trait instance
given by a \li+TraitRef+ (see §~\ref{sec:reconstruct})
together with the name of the method which is actually called.


We end our overview of LLBC here. The omitted parts of the AST follow a similar logic:
we attempt to factor out definitions, store as much information as
we can, including metadata, and make all information explicit.

\subsection{Interacting with the Rust Ecosystem}
The build system of Rust, \li+cargo+, takes care of fetching dependencies, at the
correct revision; building them recursively; and finally, invoking \li+rustc+ on the
current crate with include and library paths for all the required dependencies.
To seamlessly integrate with existing Rust projects, Charon therefore directly reuses
the \li+cargo+ infrastructure.

To do so, the \li+charon+ executable first invokes \li+cargo+; thanks to a special
environment variable, \li+cargo+ can be made to either call vanilla \li+rustc+
(for dependency analysis), or our own variant of \li+rustc+, dubbed \li+charon-driver+,
which we instrumented with additional hooks into the compiler.
When run, \li+charon-driver+ drives \li+rustc+, replacing the final compilation step to
LLVM with a program analysis step, i.e., the Charon framework.
From a user perspective, all these low-level implementation details are hidden, and
all is needed is to call the \li+charon+ executable from the root of the project (a.k.a.
crate).

At the difference of \li+cargo build+, which produces an executable,
calling \li+charon+ produces a \li+.(u)llbc+ file containing a straightforward serialization
(currently in JSON) of the (U)LLBC.
Any further analyses are to be performed off of those files;
should the analysis itself be written in Rust, we engineered Charon so that
the part that links against \li+rustc+ lives in a separate crate from the rest
(cleanups, CFG and AST representations, serialization and deserialization),
avoiding the need for analysis tools to include the whole compiler toolchain.

We can also auto-generate (U)LLBC type declarations and deserializers for other languages
beyond Rust; right now, we provide \li+charon-ml+, an
OCaml library that can read back \li+.(u)llbc+ files.
OCaml is particularly-well suited to AST manipulations, and comes
with convenient facilities, such as automatic visitors generation, which
accelerates development of analysis tools.
Adding support for a different language
would only require a modicum of work.

\subsection{Implementing Charon}

Our toolchain is made up of two parts: the Charon codebase itself,
totaling 18kLoC excluding whitespace and comments, which constructs (U)LLBC and performs additional transformation passes,
and the hax-frontend-exporter crate, developed in collaboration with the hax project~\cite{hax}
and totaling 9kLoC, which directly consumes the output of \li+rustc+ and performs most queries, as
well as our constant simplification and custom trait resolution, which operate on both MIR
and THIR, and which we believe to be useful independently of Charon as a companion to the
compiler and thus package separately.
The effort to write this whole project spanned 2 person-years.

\section{Case Studies}
\label{sec:case-studies}


\paragraph{Static Analyses.}

We implemented two static analyses on top of Charon.
First, we ported Rudra, a recent static analyzer that detects potential memory safety
issues in unsafe Rust programs by looking for a set of well-identified bug
patterns~\cite{rudra}, so that it uses Charon rather than directly interacting with \li+rustc+.
The port took us one day of work, confirming in particular that ULLBC provides all the
needed information; to validate the analysis, we reran our port of Rudra on
versions of crates used in the original paper's evaluation.  Several of these crates do
not compile anymore, due to the Rust ecosystem evolving and the projects not including
\li+.lock+ files, but we were nevertheless able to analyze 6 of the most popular
crates considered in the initial paper, reidentifying vulnerabilities previously
discovered.

We also implemented an analyzer to detect constant-time violations in cryptographic code
through a flow-, field-, and context-sensitive taint analysis operating on LLBC. We
ran our analysis on implementations from several cryptographic
crates, including RustCrypto~\cite{rustcrypto}, a port to
safe Rust of the formally verified \haclstar library~\cite{haclxn,hacl-rs},
and an
implementation of the recently standardized post-quantum
ML-KEM~\cite{ml-kem-standard} cryptographic primitive
in the libcrux formally verified cryptographic library~\cite{libcrux},
totalling 88k LoC.
Our taint analysis successfully shows that the considered implementations do
not suffer from constant-time violations, as expected from widely-used or
verified libraries, and rediscovers the KyberSlash timing attack in an
earlier, unverified version of ML-KEM~\cite{kyberslash,kyberslash-blog}.

\paragraph{Deductive Verification.}

Aeneas~\cite{ho2022aeneas,ho2024sound} is a framework for verifying safe Rust
programs, which works by generating models of Rust programs which are exported to a range
of theorem provers. Aeneas relies on Charon to obtain the LLBC code and was
the original motivation for implementing Charon; after realizing that many other
tools were each reimplementing their own logic for
interacting with \li+rustc+,
we felt strongly that
packaging and releasing a reusable component for this task would help current
and future Rust tool authors.

\paragraph{Transpilation to C.}

Eurydice\footnote{\url{https://github.com/AeneasVerif/eurydice}} is a transpiler from Rust to C whose main motivation is to allow engineers to
develop new code in Rust while still being able to deliver C code for legacy reasons.
It is made up of about 5000 lines of OCaml code, including whitespace and
comments, and consumes Rust code via Charon, translating LLBC into C.
Eurydice particularly benefits from LLBC's design: the AST is small and structured, simplifying the translation, and
move and copy operations, needed to correctly match the C semantics, are explicit.

\paragraph{Running Charon over Charon.}
\label{sec:charononcharon}
We mentioned earlier that Charon exposes its (U)LLBC representations in a reusable format
for use in other languages; the Charon codebase itself maintains an OCaml library
to manipulate this format, used in Eurydice and Aeneas.
Rather than author this library manually, we instead run Charon on itself to inspect the
definitions of the LLBC and ULLBC type definitions written in Rust, and output
appropriate OCaml type definitions, visitors and deserializers to read and manipulate
(U)LLBC.

\paragraph{Third-Party Uses of Charon.}
While Charon is still recent, we can already report third-party uses of the
framework. The Kani model checker~\cite{kani} recently added a backend that generates ULLBC so as
to leverage Charon's control-flow reconstruction pass. RaRust~\cite{rarust-github} is a linear
resource bound analysis for Rust which uses (an earlier version of) Charon to
retrieve the LLBC of Rust programs.

\section{Related Work}

To the best of our knowledge, the only other project that aims to provide a view
over MIR suitable for a variety of tooling is Stable-MIR~\cite{stable-mir}.
Much like Charon, Stable-MIR
has its own representation of important Rust
constructs and features strongly-typed identifiers.
Unlike Charon, Stable-MIR is not in itself a standalone \li`rustc` driver;
it is instead closer to a toolkit to write \li`rustc` drivers.
While it does considerably simplify the interactions
with the compiler by providing appropriate methods instead
of out-of-band queries and hiding away the driver details,
it is by design a thin wrapper over compiler internals.
As such, it does not intend to clean up or reconstruct the CFG into an AST
nor does it try to simplify constants or resolve traits as we do.
Both projects have similar goals however,
and may fruitfully collaborate in the future.

The Charon project is collaborating with other projects that have similar needs.
For instance, hax~\cite{hax} and Charon share important components including
their trait resolution system and
simplification of constants,
while Kani~\cite{kani} recently added a backend that generates ULLBC so as
to leverage Charon's control-flow reconstruction pass.
Generally, for historical reasons, many of the existing verifiers and/or
Rust-based tools maintain their own equivalent functionality, in a less general
form and more tailored to their own needs.
We hope for the adoption of Charon to
continue and plan to formalize a roadmap, governance model, and community, to
make sure more tools and clients can rely on Charon.

Beyond Rust, several efforts have proposed intermediate representations
that other tools can either target or consume.
Most famously, the LLVM toolchain provides a common substrate for many source
languages, usable for many analyses~\cite{heapster,gurfinkel2021abstract,grech2015static,zhao2012formalizing}.
While LLVM was initially intended as a compilation framework~\cite{llvm}, it
now includes building blocks that significantly reduce the effort needed to
develop new analyses, e.g., libraries for dominators or alias analysis; we intend
to provide similar features for (U)LLBC.
Runtimes for managed languages like the JVM and the .NET CLR have
been targeted by many languages (e.g., Scala, Clojure,
F\#), and have in turn created a foundation for general-purpose analyses (e.g.,
CLR Profiler, Abstract Interpretation for Java~\cite{marx2024abstract}).
Similarly to Charon, Emscripten, the LLVM backend for WASM, also performs some cleanups to
reconstruct structured control flow with the Relooper algorithm, so as to
target the more structured semantics of WASM~\cite{zakai11emscripten}.

\begin{credits}
\subsubsection{\ackname}
This work received funding from the France 2030 programs managed by the French National Research Agency
under grant agreements ANR-22-PTCC-0001 and ANR-22-PETQ-0008 PQ-TLS.
\end{credits}

%
%
%
\bibliographystyle{splncs04}
\bibliography{conferences,cited}

\begin{thebibliography}{10}
\providecommand{\url}[1]{\texttt{#1}}
\providecommand{\urlprefix}{URL }
\providecommand{\doi}[1]{https://doi.org/#1}

\bibitem{thirbugs}
{Github tracking issue for bugs fixed by the MIR borrow checker or NLL}.
  \url{https://github.com/rust-lang/rust/issues/47366}

\bibitem{prusti}
Astrauskas, V., M\"uller, P., Poli, F., Summers, A.J.: Leveraging {R}ust types
  for modular specification and verification. In: Proceedings of the ACM
  SIGPLAN Conference on Object-Oriented Programming, Systems, Languages and
  Applications (OOPSLA) (2019)

\bibitem{gillian-rust}
Ayoun, S.{\'E}., Denis, X., Maksimovi{\'c}, P., Gardner, P.: A hybrid approach
  to semi-automated rust verification. arXiv preprint arXiv:2403.15122  (2024)

\bibitem{rudra}
Bae, Y., Kim, Y., Askar, A., Lim, J., Kim, T.: Rudra: finding memory safety
  bugs in rust at the ecosystem scale. In: Proceedings of the ACM SIGOPS 28th
  Symposium on Operating Systems Principles. pp. 84--99 (2021)

\bibitem{kyberslash}
Bernstein, D.J., Bhargavan, K., Bhasin, S., Chattopadhyay, A., Chia, T.K.,
  Kannwischer, M.J., Kiefer, F., Paiva, T., Ravi, P., Tamvada, G.:
  {KyberSlash}: Exploiting secret-dependent division timings in kyber
  implementations. Cryptology {ePrint} Archive, Paper 2024/1049 (2024),
  \url{https://eprint.iacr.org/2024/1049}

\bibitem{hax}
Bhargavan, K., Franceschino, L., Hansen, L.L., Kiefer, F., Schneider-Bensch,
  J., Spitters, B.: {Hax - Enabling High Assurance Cryptographic Software}.
  RustVerify (2024),
  \url{https://github.com/hacspec/hacspec.github.io/blob/master/RustVerify24.pdf}

\bibitem{kyberslash-blog}
Bhargavan, K., Kiefer, F., Tamvada, G.: Verified {ML-KEM} (kyber) in rust.
  \url{https://cryspen.com/post/ml-kem-implementation/}

\bibitem{charon-github}
{Charon Team}: Charon github repository (2025),
  \url{https://github.com/AeneasVerif/charon}

\bibitem{charon-artifact}
{Charon Team}: Charon zenodo artifact (2025),
  \url{https://zenodo.org/records/15314373}

\bibitem{rustlinux}
Cook, K.: { [GIT PULL] Rust introduction for v6.1-rc1.}
  \url{https://lore.kernel.org/lkml/202210010816.1317F2C@keescook/}

\bibitem{creusot}
Denis, X., Jourdan, J.H., March{\'e}, C.: Creusot: a foundry for the deductive
  verification of rust programs. In: International Conference on Formal
  Engineering Methods. pp. 90--105. Springer (2022)

\bibitem{refinedrust}
G{\"a}her, L., Sammler, M., Jung, R., Krebbers, R., Dreyer, D.: {RefinedRust: A
  type system for high-assurance verification of Rust programs}. Proceedings of
  the ACM on Programming Languages  \textbf{8}(PLDI),  1115--1139 (2024)

\bibitem{grech2015static}
Grech, N., Georgiou, K., Pallister, J., Kerrison, S., Morse, J., Eder, K.:
  Static analysis of energy consumption for llvm ir programs. In: Proceedings
  of the 18th International Workshop on Software and Compilers for Embedded
  Systems. pp. 12--21 (2015)

\bibitem{heapster}
Gritti, F., Pagani, F., Grishchenko, I., Dresel, L., Redini, N., Kruegel, C.,
  Vigna, G.: Heapster: Analyzing the security of dynamic allocators for
  monolithic firmware images. In: In Proceedings of the IEEE Symposium on
  Security \& Privacy (S\&P) (May 2022)

\bibitem{gurfinkel2021abstract}
Gurfinkel, A., Navas, J.A.: Abstract interpretation of llvm with a region-based
  memory model. In: International Workshop on Numerical Software Verification.
  pp. 122--144. Springer (2021)

\bibitem{ho2024sound}
Ho, S., Fromherz, A., Protzenko, J.: Sound borrow-checking for {Rust} via
  symbolic semantics. Proceedings of the ACM on Programming Languages
  \textbf{8}(ICFP),  426--454 (2024)

\bibitem{ho2022aeneas}
Ho, S., Protzenko, J.: Aeneas: {Rust} verification by functional translation.
  Proceedings of the ACM on Programming Languages  \textbf{6}(ICFP),  711--741
  (2022). \doi{10.1145/3547647}

\bibitem{rustbelt}
Jung, R., Jourdan, J.H., Krebbers, R., Dreyer, D.: Rustbelt: Securing the
  foundations of the {Rust} programming language. In: Proceedings of the ACM
  Symposium on Principles of Programming Languages (POPL) (2018)

\bibitem{kani}
{Kani Contributors}: {The Kani Rust Verified}.
  \url{https://github.com/model-checking/kani}

\bibitem{llvm}
Lattner, C., Adve, V.: {LLVM}: A compilation framework for lifelong program
  analysis \& transformation. In: Proceedings of the International Symposium on
  Code Generation and Optimization (CGO) (2004)

\bibitem{verus}
Lattuada, A., Hance, T., Cho, C., Brun, M., Subasinghe, I., Zhou, Y., Howell,
  J., Parno, B., Hawblitzel, C.: Verus: Verifying {Rust} programs using linear
  ghost types. In: Proceedings of the ACM SIGPLAN Conference on Object-Oriented
  Programming, Systems, Languages and Applications (OOPSLA) (2023).
  \doi{10.1145/3586037}

\bibitem{libcrux}
{libcrux Contributors}: libcrux - the formally verified crypto library.
  \url{https://github.com/cryspen/libcrux/}

\bibitem{marx2024abstract}
Marx, S., Erdweg, S.: Abstract interpretation of java bytecode in sturdy. In:
  Proceedings of the 26th ACM International Workshop on Formal Techniques for
  Java-like Programs. pp. 17--22 (2024)

\bibitem{mirBlog}
{Matsakis, Niko}: {Introducing MIR}.
  \url{https://blog.rust-lang.org/2016/04/19/MIR.html}

\bibitem{hacspec}
Merigoux, D., Kiefer, F., Bhargavan, K.: {Hacspec: succinct, executable,
  verifiable specifications for high-assurance cryptography embedded in Rust}.
  Technical report, {Inria} (Mar 2021),
  \url{https://inria.hal.science/hal-03176482}

\bibitem{miri}
{Miri Contributors}: {Miri, an Undefined Behavior detection tool for Rust}.
  \url{https://github.com/rust-lang/miri}

\bibitem{nsa2022safelang}
{National Security Agency}: Software memory safety.
  \url{https://media.defense.gov/2022/Nov/10/2003112742/-1/-1/0/CSI_SOFTWARE_MEMORY_SAFETY.PDF}
  (2022)

\bibitem{ml-kem-standard}
{NIST}: Module-lattice-based key-encapsulation mechanism standard.
  \url{https://csrc.nist.gov/pubs/fips/203/final} (2024)

\bibitem{yuga2023}
Nitin, V., Mulhern, A., Arora, S., Ray, B.: Yuga: Automatically detecting
  lifetime annotation bugs in the rust language (2023),
  \url{https://arxiv.org/abs/2310.08507}

\bibitem{yuga}
Nitin, V., Mulhern, A., Arora, S., Ray, B.: Yuga: Automatically detecting
  lifetime annotation bugs in the rust language. IEEE Transactions on Software
  Engineering  (2024)

\bibitem{peterson1973capabilities}
Peterson, W.W., Kasami, T., Tokura, N.: On the capabilities of while, repeat,
  and exit statements. Communications of the ACM  \textbf{16}(8),  503--512
  (1973)

\bibitem{haclxn}
Polubelova, M., Bhargavan, K., Protzenko, J., Beurdouche, B., Fromherz, A.,
  Kulatova, N., Zanella-B\'{e}guelin, S.: {HACLxN}: Verified generic {SIMD}
  crypto (for all your favourite platforms). In: Proceedings of the ACM
  Conference on Computer and Communications Security (CCS) (2020)

\bibitem{ramsey2022beyond}
Ramsey, N.: Beyond relooper: recursive translation of unstructured control flow
  to structured control flow (functional pearl). Proceedings of the ACM on
  Programming Languages  \textbf{6}(ICFP),  1--22 (2022)

\bibitem{mirpasses}
{Rust Compiler Team}: {Implementors of MirPass}.
  \url{https://doc.rust-lang.org/nightly/nightly-rustc/rustc_mir_transform/pass_manager/trait.MirPass.html#implementors}
  (2024)

\bibitem{rustcrypto}
{RustCrypto maintainers}: {Rust Crypto} - cryptographic algorithms written in
  pure rust. \url{https://github.com/RustCrypto}

\bibitem{stable-mir}
{Stable MIR contributors}: {Stable MIR: Define a compiler intermediate
  representation usable by external tools}.
  \url{https://github.com/rust-lang/project-stable-mir}

\bibitem{stackoverflow2023rust}
{StackOverflow}: 2023 developer survey.
  \url{https://survey.stackoverflow.co/2023/\#section-admired-and-desired-programming-scripting-and-markup-languages}
  (2023)

\bibitem{stateright}
{Stateright Contributors}: {Stateright, a model-checker for implementing
  distributed systems}. \url{https://github.com/stateright/stateright}

\bibitem{hacl-rs}
{The HACL* Team}: {A preliminary version of HACL* extracted to *safe* Rust}.
  \url{https://github.com/hacl-star/hacl-star/pull/918}

\bibitem{rarust-github}
{The RaRust Development Team}: Rarust (2025),
  \url{https://github.com/Mepy/rarust-oopsla25}

\bibitem{microsoft2022rust}
{The Register}: In {Rust} we trust: {Microsoft Azure CTO} shuns {C} and {C++}.
  \url{https://www.theregister.com/2022/09/20/rust_microsoft_c/} (2022)

\bibitem{rustinwindows}
{The Register}: Microsoft is busy rewriting core {Windows} code in memory-safe
  {Rust}. \url{https://www.theregister.com/2023/04/27/microsoft_windows_rust/}
  (2023)

\bibitem{whitehouse2024}
{The White House}: {Back to the Building Blocks: a Path Toward Secure and
  Measurable Software}.
  \url{https://www.whitehouse.gov/wp-content/uploads/2024/02/Final-ONCD-Technical-Report.pdf}

\bibitem{github2023rust}
Verdi, S.: Why rust is the most admired language among developers.
  \url{https://github.blog/2023-08-30-why-rust-is-the-most-admired-language-among-developers/}
  (2023)

\bibitem{zakai11emscripten}
Zakai, A.: Emscripten: an llvm-to-javascript compiler. In: Proceedings of the
  ACM SIGPLAN Conference on Object-Oriented Programming, Systems, Languages and
  Applications (OOPSLA) (2011)

\bibitem{zhao2012formalizing}
Zhao, J., Nagarakatte, S., Martin, M.M., Zdancewic, S.: Formalizing the llvm
  intermediate representation for verified program transformations. In:
  Proceedings of the 39th annual ACM SIGPLAN-SIGACT symposium on Principles of
  programming languages. pp. 427--440 (2012)

\end{thebibliography}

\end{document}